 \documentclass[final,3p,times,twocolumn]{elsarticle}
 \usepackage{graphicx}
\usepackage{amssymb,amsmath}
\journal{Journal of Theoretical Biology}

\begin{document}

\begin{frontmatter}



\title{Exact Results for Amplitude Spectra of Fitness Landscapes}


\author{Johannes Neidhart}
\author{Ivan G. Szendro}
\author{Joachim Krug}

\address{Institute of Theoretical Physics, University of Cologne, Z\"ulpicher Stra{\ss}e 77, 50937 Cologne, Germany}

\begin{abstract}
Starting from fitness correlation functions, we calculate exact
expressions for the amplitude spectra of fitness landscapes as defined
by P.F. Stadler [J.\ Math.\ Chem.\ \textbf{20}, 1 (1996)]
%E. Weinberger [Biol. Cybern. \textbf{65}, 321 (1991)] 
for common
landscape models, including Kauffman's $NK$-model, rough Mount Fuji
landscapes and general linear superpositions of such landscapes. We
further show that correlations decaying exponentially with the Hamming
distance yield exponentially decaying spectra similar to those reported
recently for a model of molecular signal transduction. Finally, we
compare our results for the model systems to the spectra of various
experimentally measured fitness landscapes. 
We claim that our analytical results should be helpful when trying to
interpret empirical data and guide the search for improved 
fitness landscape models.  
\end{abstract}


\end{frontmatter}

% \linenumbers

\section{Introduction}
\label{introduction}
In evolutionary processes, populations acquire changes to their gene content
by mutational or recombinational events during reproduction. If 
those changes improve the adaptation of the organism to its environment, individuals carrying
the modified genome have a better chance to survive and leave more offspring in the next
generation. Through the interplay of repeated mutation and selection,
the genetic structure of the population evolves and beneficial alleles increase in frequency.  
In a constant environment the population may thus end up in a well
adapted state, where beneficial mutations are rare or entirely absent and only combinations of several
mutations can further increase fitness.

To describe this kind of process, Sewall Wright introduced the notion
of a fitness landscape \cite{Wright1932}.
Here, the genotype is encoded by the coordinates of some suitable space and the degree of adaptation or reproductive success is modeled as a
real number, called fitness, which is identified with the height of
the landscape above the corresponding genotype. The evolutionary process of
repeated mutation and selection is thus depicted as a hill climbing process. Mutations lead to the exploration of
new genotypes and selection forces populations to move preferentially to genotypes with larger fitness.
If more than one mutation is necessary to increase fitness, the
population has reached a local fitness peak. Note that
some caution is necessary when applying this picture, as the way in which genotypes are connected to one another does not
correspond to the topology of a low-dimensional Euclidean space but is more appropriately described by a graph or
network (see below). The underlying structure is well known from other areas of
science, such as spin glasses in statistical physics \cite{binderyoung,parisi} and optimization problems in computer science \cite{gareyjohnson}.

The concept of fitness landscapes has been very fruitful for the understanding of evolutionary processes.
While earlier work in this field has been largely theoretical and
computational, in recent years an increasing amount of
experimental fitness data for mutational landscapes has become
available
\cite{Lunzer2005,Weinreich2006,Poelwijk2007,OMaille2008,Lozovsky2009,Brown2010,Hall2010,DaSilva2010,Costanzo2011,Chou2011,Khan2011,Tan2011},
see ref.~\cite{Szendro2012} for a review. Analysis of such  
data sets provides us with the possibility of a better understanding of the biological mechanisms that shape fitness landscapes 
and helps us to build better models. Thus, identifying properties 
of fitness landscapes that yield relevant information on evolution is an 
important task. 

One such property that has attracted
considerable interest is epistasis \cite{deVisser2011}.
Epistasis implies that the change in fitness that is caused by a specific mutation 
depends on the configurations at other loci, or groups of loci, in the genome. 
In other words, epistasis is the interaction between different loci in
their effect on fitness. Interactions that only affect the strength of the
mutational effect are referred to as magnitude epistasis, while
interactions that change a mutation from beneficial to deleterious
or vice versa are referred to as sign epistasis \cite{Weinreich2005}. In the absence of sign epistasis, the fitness landscape contains 
only a single peak and fitness values fall off monotonically with distance to that peak. If sign epistasis is present, 
the landscape can present several peaks and valleys, which has
important implications for the mutational accessibility of 
the different genotypes \cite{Poelwijk2007,jaspjoa,Franke2012} and
shortens the path to the next fitness optimum
\cite{orr1,joyceorr,Poelwijk2010,krugneid,Crona2013}. 
Thus the absence of sign epistasis implies a smooth landscape, while landscapes with
sign epistasis are rugged.  

Beyond the question of the presence of epistasis, one would like to be able to make more detailed statements about {\it how much} of it is present or {\it in which way} epistasis is realized in the landscape. A very helpful tool to answer
these kinds of questions is the Fourier decomposition of fitness landscapes introduced in ref.~\cite{Weinberger1991}. This decomposition makes
use of graph theory to expand the landscape into components that correspond to interactions between 
loci. The coefficients of the decomposition corresponding to
interactions between a given number of loci can be combined to yield
the \textit{amplitude spectrum}. Calculating amplitude spectra
numerically for data obtained from models or experiments is
straightforward in principle, but so far only a small part of the
information contained in the spectra is actually used. 
To improve this situation, it is important to understand how biologically
meaningful features of a fitness landscape are reflected in its
amplitude spectrum.

In this paper, we take a first step in this direction by analytically calculating spectra for some 
of the most popular landscape models: the $NK$-model introduced by Kauffman \cite{kauffwein,kauffman}, two versions of the rough Mount Fuji 
(RMF) model \cite{jaspjoa,aita}, and a generic model with correlations that decay exponentially with distance on the 
landscape. Thanks to the linearity of the amplitude decomposition,
linear superpositions of these landscapes can also be treated. We
calculate the spectra by exploiting their connection to fitness
correlation functions originally established in ref.~\cite{stadler3}. Moreover, we compare some experimentally obtained spectra to the predictions of the models to see what features 
can be explained by these models and which can not. In the next
section we begin by introducing the definitions of fitness landscapes and 
their amplitude spectra on more rigorous mathematical grounds.

\section{Fitness landscapes and their amplitude spectra}

\subsection{Sequence space and epistasis}

   \begin{figure*}[htb]
    \begin{center}
    \includegraphics[width=.8\textwidth]{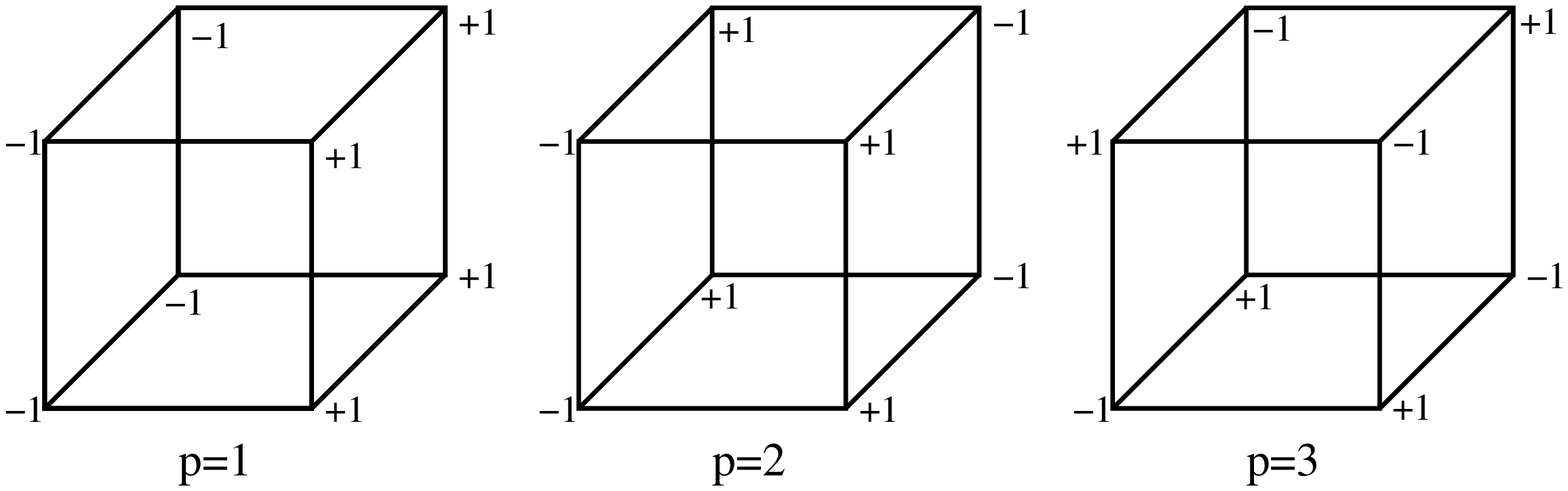}
    \caption{Illustration of the eigenfunctions $2^{N/2} \phi_{i_1,\dots,i_p}$ of the graph Laplacian for the binary
      hypercube with $N=3$ and $p=1,2,3$. Similar to the usual Fourier decomposition
      on spaces such as $\mathbb{Z}^n$ or $\mathbb{R}^n$,
      eigenfunctions of higher order vary more rapidly.}
    \label{fig:eigenfct}
    \end{center}
    \end{figure*}

    On the molecular level
    the genotype of an organism is encoded in a
    sequence of letters taken from the alphabet $\mathfrak A = \{T,C,G,A\}$ of
    nucleotide base pairs with cardinality $|\mathfrak A|=4$. Point mutations replace
    single letters by others, altering the sequence and therefore the
    properties of the organism. A
    similar description applies to the space of proteins, where
    the cardinality of the encoding alphabet equals the number of
    amino acids \cite{MaynardSmith1970}. By contrast, in the context of classical 
    genetics the units making up the genotype are genes occurring in different variants 
(alleles), which again can be described as letters in some alphabet \cite{Wright1932}. This provides a coarse-grained
view of the genome in which also complex mutational events are represented by replacing one allele by another. 

    For simplicity, fitness landscapes are often defined on sequences
    comprised of elements of a \textit{binary} alphabet $\mathfrak A^B$, where
    a common choice is $\mathfrak A^B=\{0,1\}$. In the present
    article we prefer the symmetric alphabet $\mathfrak A^B=\{-1,1\}$ for
    mathematical convenience \cite{Neher2011}. Referring to the discussion in the preceding paragraph, 
    we emphasize that the elements of
    the binary alphabet do not generally stand for bases or encoded
    proteins but rather indicate whether a particular 
    mutation is present in a gene or not \cite{Szendro2012}. Therefore the restriction to 
    single changes in the sequence does not imply that the treatment is
    limited to point mutations. 

    All possible sequences of a given length $N$ constructed from the
    alphabet $\mathfrak A$ with cardinality $|\mathfrak A|=\kappa$ form a metric space called the
    \textit{Hamming space} $\mathbb H_\kappa^N$. It can be expressed as  $\mathbb H_\kappa^N = (\mathcal K_\kappa)^N = \mathcal K_\kappa\otimes\dots\otimes
    \mathcal K_\kappa$, where
    $\otimes$ denotes the
    Cartesian product and $\mathcal K_\kappa$ is the complete
    graph with $\kappa$ nodes.     
    For a binary alphabet the $\mathbb H_2^N$ are hypercubes. Their metric is called the
    \textit{Hamming distance},
    \begin{align}
        d:\mathbb H_2^N \times \mathbb H_2^N &\to \mathbb N\cup \{0\}\nonumber\\
        (\sigma, \sigma')&\mapsto \sum_{i=1}^N (1-\delta_{\sigma_i, \sigma'_i })\,
    \end{align}
which equals the number of single mutational steps required to
transform one sequence into the other. 
    To quantify the degree of adaptation or reproductive success of an
    organism carrying the genotype $\sigma$, a real number $F$ called 
    fitness is assigned to the corresponding
    sequence according to
    \begin{align}
        F:\mathbb H_2^N &\to \mathbb R\nonumber\\
        \sigma &\mapsto F(\sigma).
    \end{align}
    To precisely define the different notions of epistasis introduced
    above, we consider two sequences 
    $\sigma, \sigma' \in \mathbb H_2^N$ with $d(\sigma,\sigma')<N$. Let $\sigma=(\sigma_1,\dots,
    \sigma_i,\dots, \sigma_N)$ and $\sigma' =
    (\sigma'_1,\dots,\sigma_i,\dots, \sigma'_N)$, and denote the
    sequences with a mutation at the $i$--th locus by $\sigma^{(i)}$
    and $\sigma'^{(i)}$, respectively, with 
    $\sigma_i^{(i)} = {\sigma'}_i^{(i)} = - \sigma_i$). If $F(\sigma)-F(\sigma^{(i)})\neq
    F(\sigma')-F(\sigma'^{(i)})$ for some $i$, the fitness landscape is called \textit{epistatic}. If
    $\text{sgn}(F(\sigma)-F(\sigma^{(i)}))=\text{sgn}(F(\sigma')-F(\sigma'^{(i)}))$ the effect is called \textit{magnitude
    epistasis}, while for $\text{sgn}(F(\sigma)-F(\sigma^{(i)}))=-\text{sgn}(F(\sigma')-F(\sigma'^{(i)}))$ it is called
    \textit{sign epistasis}. Furthermore, the landscape is said to
    contain \textit{reciprocal sign epistasis} if there are pairs of
    mutations such that
    $-\text{sgn}(F(\sigma)-F(\sigma^{(i,j)}))=\text{sgn}(F(\sigma)-F(\sigma^{(i)}))=\text{sgn}(F(\sigma)-F(\sigma^{(j)}))$,
    with $\sigma^{(i,j)}$ denoting the sequence mutated at loci $i$
    and $j$ \cite{Poelwijk2007}. A landscape with sign epistasis is
    said to be \textit{rugged}, while landscapes containing no
    epistasis or only magnitude epistasis are called \textit{smooth}. 
Non-epistatic landscapes are also called \textit{additive}, as here the individual effects of mutations add up independently. 

The presence of sign epistasis severely limits which paths on the
landscape are accessible to evolution
\cite{Poelwijk2007,Weinreich2005,jaspjoa}. Landscapes that display reciprocal
sign epistasis may contain several local fitness maxima
\cite{Poelwijk2010}, while those that do not have a single
maximum. The existence of reciprocal sign epistasis is a necessary but
not sufficient condition for the existence of multiple
maxima. For an example of a sufficient condition for multiple maxima
based on local properties of the landscape see \cite{Crona2013}.
   
\subsection{Fourier decomposition}

    The \textit{adjacency matrix} $\mathcal A$ of the Hamming space encodes the
    neighborhood relations between sequences, and is defined as
    \begin{align}
        \mathcal A_{\sigma, \sigma'}=\begin{cases} 
					1, & d(\sigma, \sigma')=1\\
                                        0, &\text{else}.
                                     \end{cases}
    \end{align}
    With $\mathbb I^m$ denoting the identity of $m\times m$ matrices,
    the graph Laplacian $\Delta$ is then defined by $\Delta=\mathcal
    A-N\mathbb I^{2^N}$, and its action on the fitness function $F$ yields 
    \begin{align}
        \Delta F(\sigma) &= \sum_{\sigma'\in\mathbb H_2^N} \mathcal A_{\sigma,\sigma'}F(\sigma')-NF(\sigma)\nonumber\\
                        &= \sum_{\stackrel{\sigma'\in\mathbb H_2^N,}{d(\sigma, \sigma')=1}} F(\sigma')-NF(\sigma).
    \end{align}
    For $\mathfrak A=\mathfrak A^B = \{-1,1 \}$ and $\sigma_i$ denoting the $i$--th element of $\sigma$, the eigenfunctions of $\Delta$ 
    are given by $\phi_{i_1,\dots, i_p}(\sigma)=2^{-\frac N 2} \sigma_{i_1}\dots\sigma_{i_p}$
    with $p\in\{1,\dots, N\}$ and $0\leq i_1\leq i_2 \dots\leq i_p\leq N$. The corresponding eigenvalues are $\lambda_p
    = -2p$ and thus the degeneracy is $\binom N p$. The set of all eigenfunctions $\phi_i(\sigma)$ forms an orthonormal  basis and the landscape can be
    expressed in terms of a decomposition, called \textit{Fourier
      expansion} \cite{Weinberger1991}, which reads
    \begin{align}
        F(\sigma) = \sum_{p=0}^N\sum_{i_1\dots i_p } a_{i_1\dots i_p}\phi_{i_1\dots i_p}(\sigma). \label{eq:ft}
    \end{align}
    See fig.\ \ref{fig:eigenfct} for the visualization of three eigenfunctions on the $N=3$ hypercube.
   While the $a_{i}$'s contain the information about the relative influence of the non-epistatic
   contributions on fitness, the higher order coefficients $a_{i_1\dots i_p}$ with $p>1$ describe the relative strength of
   the contributions of $p$--tupels of interacting loci. The zero
   order coefficient $a_0$ is proportional to the mean fitness of the
   landscape, $$a_0 = 2^{-\frac N 2}\sum_{\sigma\in \mathbb
   H^2_N}F(\sigma),$$ where the prefactor reflects the normalization of the $\phi_i$. 

The amplitude spectrum quantifies the relative contributions of the complete sets of $p$--tupels to the epistatic interactions. Following ref.~\cite{stadler3}, we consider \textit{random
     field models} of fitness landscapes where individual instances of
   the ensemble
   (\textit{realizations}) are constructed from random variables
   according to some specified rule (see sects.~\ref{nkmodel} and \ref{rmfmodel}), and define
   amplitude spectra as averages over the realizations. Two kinds of
   averages appear: averaging over realizations at a constant point in $\mathbb H_2^N$, and
   \textit{spatially} averaging over all points in $\mathbb H_2^N$. Here and in the following
    angular brackets $\langle\ldots\rangle$ denote averaging over the realizations
    of the landscape, while an overbar denotes a spatial average over $\mathbb
    H_2^N$, as for example in 
$$\overline{F}=2^{-N}\sum_\sigma F(\sigma).$$

    For the definition of the amplitude spectrum the two types of
   averages need to be distinguished. The first one reads 
    \begin{align}
      B_p = \left \langle\frac{\sum_{i_1\dots i_p}|a_{i_1\dots i_p}|^2}{\sum_{q\neq 0} \sum_{i_1\dots i_q}|a_{i_1\dots i_q}|^2}\right\rangle, 
    \end{align}
    for $p> 0$ and $B_0=0$. For an additive landscape $B_1=1$ and
    $B_\mathrm{sum}=\sum_{i>1}B_i=0$ while for a landscape with
    epistasis $B_\mathrm{sum}>0$. In \cite{Szendro2012}
    $B_\mathrm{sum}$ was used as a quantifier for the amount of
    epistasis found in empirical fitness landscapes. Note that the values of 
    $B_\mathrm{sum}$ for different landscapes are contrastable because of the normalization $\sum_{p> 0}B_p=1$.
    
Another way to define the amplitude spectrum is through
    \begin{align}
        \tilde B_p = \frac{b_p}{b_0 + \sum_{q\neq 0} b_q},
    \end{align}
    with $b_p= \sum_{i_1\dots i_p}\langle |a_{i_1\dots i_p}|^2\rangle$
    for all $p\geq 1$. The zero order coefficient
    $b_0$ is not defined in terms of the Fourier coefficients $a_i$,
    but is proportional to the  mean covariance,
    \begin{align}
         b_0=2^{-N}\sum_{\sigma,\sigma'\in \mathbb H_2^N}\left[ \langle F(\sigma)F(\sigma')\rangle -\langle F(\sigma)\rangle
        \langle F(\sigma')\rangle\right],
    \end{align}
   as defined\footnote{Note that the prefactor of
     $b_0$ given in \cite{stadler3} appears to be incorrect.} in \cite{stadler3}. 
    The main difference
    between the $\tilde B_p$ and the $B_p$ consists in whether averaging is performed separately on the terms
    in the fraction or on the fraction as a whole. As it is often easier to calculate a fraction of averages than an
    average of a fraction, the present work concentrates on  the $\tilde
    B_p$. While the $\tilde B_p$ are not generally normalized, $\sum_{p>
      0}\tilde B_p\neq 1$, a normalized amplitude spectrum can
    easily be constructed through
    \begin{align}
    B_p^*=\frac{\tilde B_p}{\sum_{q>0} \tilde B_q} = \frac{b_p}{\sum_{q> 0} b_q}.\label{eq:bstar}
    \end{align}
   
\subsection{Relation to fitness correlations}

    In ref.~\cite{stadler3} it was shown that the differently averaged
    spectra are related to different types of fitness correlation
    functions.
    The \textit{direct correlation function}  is defined for all sequences of a given Hamming distance $d$ as
    \begin{align}
        \rho_d = \frac{1}{\binom{N}{d}2^{N}}\sum_{\stackrel{\sigma, \sigma'\in\mathbb H_2^N}{d(\sigma, \sigma')=d}} \frac{(F(\sigma)-\overline{F})(F(\sigma^\prime)- \overline{F} )}{ \overline{F^2} -\overline{F} ^2}.
    \end{align}
    This correlation function is linked to the normalized amplitude spectrum, $B_p$, according to
    \begin{align}
        \langle \rho_d \rangle = \sum_{p\geq 0}B_p \, \omega_p(d)\label{eq:directcorr}
    \end{align}
    where the $\omega_p$ are orthogonal functions depending on the underlying graph structure \cite{stadler3}.
    On the other hand, the \textit{autocorrelation function} $R_d$
    defined as\footnote{This is a slight variation of the autocorrelation function given in ref.~\cite{stadler3}. The
    original definition is restricted to landscape models fulfilling $\langle F(\sigma)\rangle=\text{const.}$, with a constant that is independent of $\sigma$. The proof of
    Theorem 5 in \cite{stadler3} can be carried out analogously for the definition (\ref{eq:autocorr}) without
    suffering from this constraint.}
    \begin{align}
        R_d= \frac{\left\langle {F(\sigma)F(\sigma')}\right\rangle_d -\left\langle  \overline{F} \right\rangle^2}{ \left\langle \overline{F^2}\right\rangle-\left\langle \overline{F} \right\rangle^2 },\label{eq:autocorr}
    \end{align}
    where $\langle\ldots\rangle_d$ denotes a \textit{simultaneous} average over all possible pairs ($\sigma,\sigma^\prime$) with
    $d(\sigma,\sigma^\prime)=d$ as well as over the realizations of the landscape, is linked to the amplitude spectrum $\tilde B_p$ according to \cite{stadler1}
    \begin{align}
        R_d = \sum_{p\geq 0 } \tilde B_p \, \omega_p(d)\label{eq:kt1}.
    \end{align}
    Again, the difference between eq.\ (\ref{eq:kt1}) and eq.\
    (\ref{eq:directcorr}) lies in how the averaging is performed.
    
For Hamming graphs $\mathbb H_\kappa^N$, the functions $\omega_p(d)$ are closely related to
the \textit{Krawtchouk polynomials} $K^{(\kappa)}_p(d)$ \cite{stadler3}. For the binary case:
$$\omega_p(d) = \binom{N}{p}^{-1} K^{(2)}_p(d),$$ where \cite{Szego,stoll}  
    \begin{align}
    K^{(2)}_p(d) = \sum_{j\geq 0} (-1)^j \binom{d}{j}\binom{N-d}{p-j}.
    \end{align}
    Unless stated otherwise, here and in the rest of the paper, binomial coefficients are understood to be defined as  
    \begin{equation}
    \binom
    N k = \begin{cases}
    \frac{N!}{k!(N-k)!} , & N\geq k \text{ and } N,k \geq 0,\\
    0,&\text{else}.
    \end{cases} \label{convention}                                                                                  
    \end{equation}

Our primary interest is in the calculation of analytical expressions of the $\tilde B_p$ for known $R_d$.
    Thus, an inversion of eq.~(\ref{eq:kt1}) is needed.
This can be achieved by exploiting the orthogonality of the Krawtchouk
polynomials with respect to the binomial distribution, which implies that    
\cite{Szego}
    \begin{equation}
%         \langle K^{(2)}_p,K^{(2)}_q\rangle =
\sum_{d\geq 0 } \binom{N}{d}K^{(2)}_p(d)K^{(2)}_q(d) =
        2^N\binom{N}{p}\delta_{pq}. 
    \end{equation}
Multiplying eq.~(\ref{eq:kt1}) by $\binom{N}{d} K^{(2)}_q(d)$ and summing over $d$
thus yields
    \begin{equation}
\sum_{d\geq 0}\sum_{p\geq 0} \binom{N}{d} \binom{N}{p}^{-1}
       \tilde B_p K^{(2)}_p(d) K^{(2)}_q(d) = 2^N \tilde B_q,
\end{equation}  
and we conclude that 
    \begin{equation}
     \tilde{B}_q=2^{-N}\sum_{d\geq 0}K^{(2)}_q(d)\binom N d R_d.\label{eq:bq}
    \end{equation}
    Now, the calculation of amplitude spectra from autocorrelation functions is possible and at least numerically
    any spectrum can be calculated from a given correlation
    function. But for some landscape models even exact analytical solutions can be obtained, as will be shown in the following
    sections.

\section{Kauffman's NK-model}
\label{nkmodel}

    \begin{figure}[htb]
    \begin{center}
    \begin{minipage}{0.9\columnwidth}
    \includegraphics[width=\columnwidth]{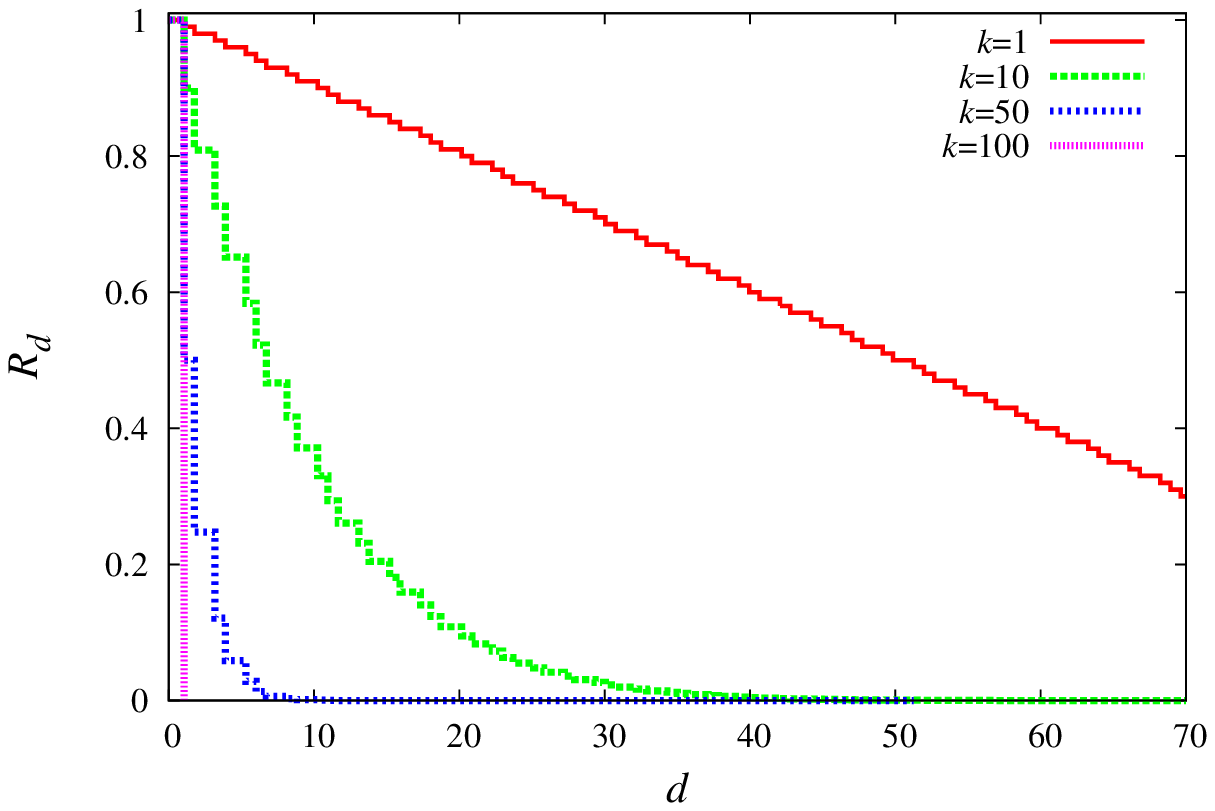}
    \end{minipage}
    \begin{minipage}{0.9\columnwidth}
    \includegraphics[width=\columnwidth]{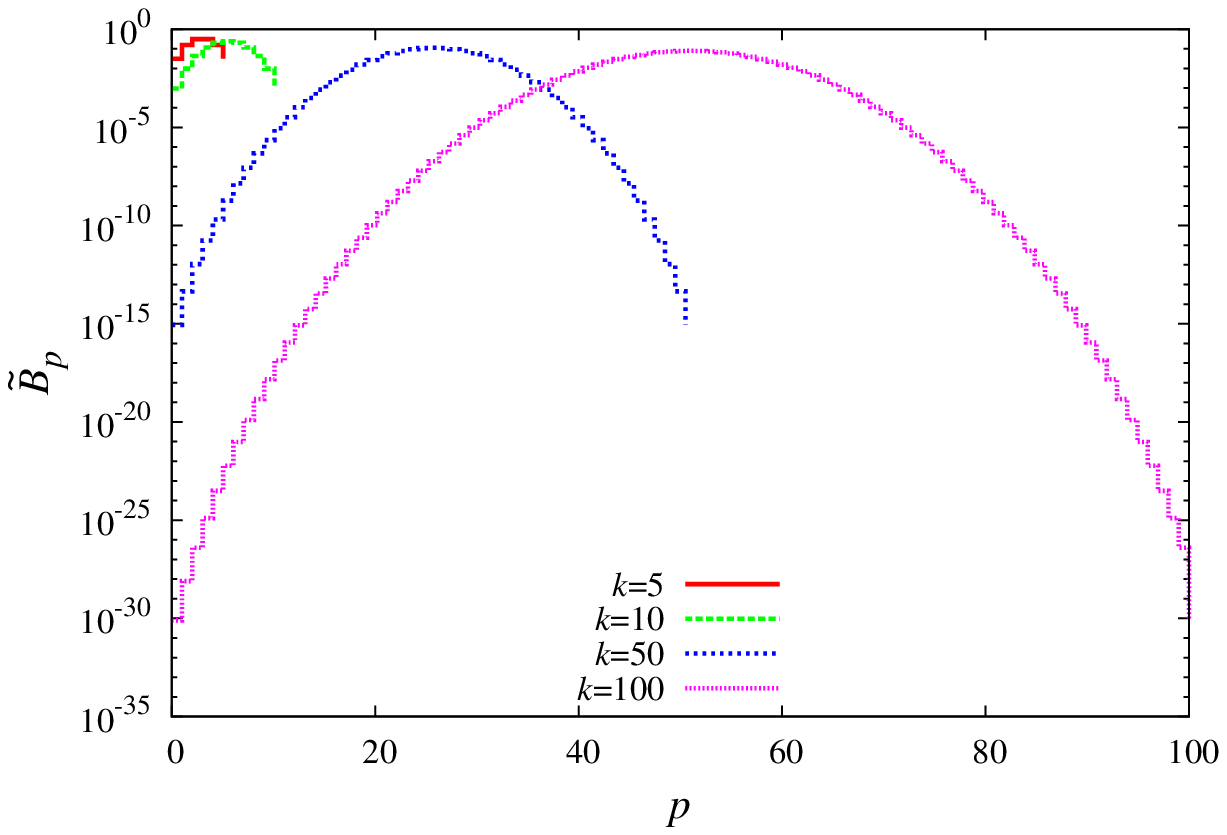}
    \end{minipage}
    \caption{The autocorrelation function (top) and the % normalized 
amplitude spectrum (bottom) for the $NK$-model with $N=100$ and different values of $k=K+1$.}
    \label{fig:autocorrbplk}
    \end{center}
    \end{figure}

    The simplest random field model of a fitness landscape is the
    \textit{House-of-Cards} (HoC) model \cite{kingman2, kaufflev}. In
    this model, the fitness values are assigned randomly to genotypes according to 
    \begin{equation}
 %       F:\mathbb H_2^N &\to \mathbb R\nonumber\\
        F:\sigma \mapsto \xi(\sigma),
    \end{equation}
    where the $\xi(\sigma)$ are independent and identically
    distributed (i.i.d.) random variables drawn from some
    distribution. Without loss of generality we assume that the $\xi$
    have vanishing mean, $\langle \xi \rangle = 0$ and finite variance
    $D=\mathrm{Var}(\xi)$. 
    The amplitude spectrum of the HoC-model is known to be $\tilde
    B_q=2^{-N}\binom{N}{q}$ \cite{stadler3},  which also follows from
    eq. (\ref{eq:bq}).
    
Although the HoC-model has been widely used for the modeling of
adaptation \cite{orr1,joyceorr,krugneid}, there is by now substantial
experimental evidence that the assumption of uncorrelated fitness values
overestimates the ruggedness of real fitness landscapes
\cite{Szendro2012,jaspjoa,miller}. It is therefore necessary to
consider more complex models, which include fitness correlations 
in a  biologically meaningful way. A prototypical model with tunable ruggedness is
Kauffman's $NK$-model \cite{kauffwein,kauffman,Welch2005}, which assumes random epistatic interactions within groups of loci of fixed size 
and fixed membership. In the classical version, each locus $i$ from a sequence of total length $N$ interacts
    with a set of  $K$ other loci $\{\sigma_{i_1},\dots,\sigma_{i_K}\}$ from the same sequence, which together with the locus $\sigma_i$ itself
constitute the $NK$\textit{-neighborhood} of locus $i$. 

To take into account more general setups, the constraint of $\sigma_i$
being a member of the $i$--th neighborhood will be relaxed here \cite{stadler3,Altenberg1997}. Thus, defining $k=K+1$, the
    $i$--th $NK$-neighborhood is the set $\{\sigma_{i_1},\dots,\sigma_{i_k}\}$.
    The fitness is assigned as follows: let $\{f_i\}$ be $N$ random
    functions with $K+1=k$ binary arguments. For each of the $2^k$
    combinations of the arguments, the $f_i(\sigma_{i_1},\dots,\sigma_{i_k})$ are chosen as i.i.d.\ random variables with variance $D$. The fitness landscape is then defined as
    \begin{equation}
%        F:\mathbb H_2^N &\to \mathbb R\nonumber\\
        F:\sigma \mapsto \frac{1}{\sqrt{N}}\sum_{i=1}^N f_i(\sigma_{i_1},\dots,\sigma_{i_k}).
    \end{equation}
    Thus, each $f_i$ is equivalent to a HoC-landscape of size $K+1=k$.
    For $K=N-1$, respectively $k=N$, the landscape is maximally rugged
    and reduces to the totally uncorrelated HoC
-model, while for $K=0$, respectively $k=1$, all fitness
    contributions are independent, and the model is fully
    additive. By changing $k$ the ruggedness of the fitness
    landscape can be tuned. 

 To complete the definition of the model, it has to be specified how
 the elements of the neighborhoods are chosen. In the most commonly
 used versions of the model, the $k$ interacting loci are either
 picked at random or taken to be adjacent along the sequence
 \cite{kauffwein,kauffman}. A third possibility is to subdivide the
 sequence into blocks of size $k$, such that within blocks
 every locus interacts with every other but blocks are mutually
 independent \cite{perelsonmacken,Orr2006}. 
Although the construction of the neighborhoods affects certain
properties of the landscapes such as the number of local fitness
maxima \cite{Weinberger1991a,Limic2004} and the evolutionary
accessibility of the global maximum \cite{Franke2012,schmiegelt}, it
turns out that the autocorrelation function does not depend on it. 
The autocorrelation function of the $NK$-model can be calculated starting from eq.\ (\ref{eq:autocorr}) and is given by \cite{Campos2002}
    \begin{align}
        R_d = \binom{N-k}{d}\binom{N}{d}^{-1},\label{RdNK}
    \end{align}
    see fig. \ref{fig:autocorrbplk}. Note that previously some incorrect expressions
    for the correlation functions have been reported in the literature
    \cite{fontana} which led to the erroneous conclusion that the
    choice of the neighborhood affects the amplitude spectra \cite{stadler3}.

    Inserting (\ref{RdNK}) into eq.\ (\ref{eq:bq}) yields
    \begin{equation}
\label{Bq_NK}
        \tilde B_q = 2^{-N}\sum_{d\geq 0}K^{(2)}_q(d) \binom{N-k}{d}.
    \end{equation}
The evaluation of this expression is somewhat technical and can be
found in Appendix A.
The final result 
    \begin{align}
  %  \tilde B_q &=2^{-k}\sum_{i\geq 0}(-1)^i(-1)^{q-i}\binom{N-k}{i}\binom{q-i-(N-i)-1}{q-i}\nonumber\\
  %  &=2^{-k}(-1)^q\sum_{i\geq 0}\binom{N-k}{i}\binom{q-N-1}{q-i}\nonumber\\
  %  &=2^{-k}(-1)^q\binom{q-k-1}{q}\nonumber\\
  %  &=2^{-k}\binom{k}{q}.
    \tilde B_q =2^{-k}\binom{k}{q} \label{eq:lkspectra}
    \end{align}
is remarkably simple (see fig.\ \ref{fig:autocorrbplk} for illustration).   
%     Thus, the $\tilde B_q$, which carry valuable information of the
%     fitness landscape, have an extraordinary simple form (see fig.\
%     \ref{fig:autocorrbplk}). 
As expected, the Fourier coefficients vanish for $q > k$ \cite{Franke2012,Drossel2001}
and the known case of the HoC-model is reproduced for
$k=N$. Moreover, the coefficients satisfy the symmetry  $\tilde
B_q=\tilde B_{k-q}$ and are maximal for $q=k/2$, as was
previously conjectured in \cite{stadler3}.

     The $NK$-model is already a very flexible model and offers many possibilities for tuning. An even more general model
     is obtained by considering \textit{superpositions} of $NK$-models, in the sense of $NK$-fitness landscapes being added independently. Let $\{F_m(\sigma)=\frac{1}{\sqrt{N}}\sum_j 
     f^{(m)}_j(\sigma_{j_1},\dots,\sigma_{j_{k^{(m)}}})\}$ be a family of $n$  $NK$-fitness
     landscapes with neighborhood sizes $k^{(m)}$, $m=1,...,n$. Then its superposition $\mathcal{F}$ is defined by 
     \begin{align}
%         :\mathbb H_2^N &\to \mathbb R\nonumber\\
         \mathcal F: \sigma&\mapsto \sum_{m=1}^n F_m(\sigma) \nonumber \\
&=\sum_{m=1}^n\frac{1}{\sqrt{N}}\sum_{j=1}^{N}
         f^{(m)}_j(\sigma_{j_1},\dots,\sigma_{j_{k^{(m)}}}).    
     \end{align}
     Since the different $NK$-landscapes $\{F_m\}$ are independent, the correlation functions are additive,
    \begin{align}
        R^{\mathcal F}_d =\frac{\sum_{m= 1}^n\binom{N-k^{(m)}}{d} \binom{N}{d}^{-1} D_m}{\sum_{j=1}^n D_j}=:\sum_{i=
        0}^N A_i\binom{N-i}{d} \binom{N}{d}^{-1}\label{linearlkmodel},
    \end{align}
    with statistical weights 
$$A_i =  \sum_{\{m| k^{(m)}=i\}}\frac{D_m}{\sum_{j=1}^n D_j},$$ where
$D_m=\mathrm{Var}(f^{(m)})$. 
The sum is over all landscapes with neighborhoods of size $i$ and contains a zeroth order term that shifts the correlation
function by a constant. 
    %The $A_j =\frac{D_j}{\sum_{i=0}^N D_i}$ can be understood as normalized variances.
    The amplitude spectrum of the superposition is thus of the form
    \begin{align}
        \tilde B^\mathcal F_p = \sum_{i\geq 0} 2^{-i}A_i\binom i p. \label{eq:lkcompo}
    \end{align}
   Note that the consistent interpretation of an empirical fitness landscape as a superposition of $NK$-landscapes requires all $A_i$ to be positive. Nevertheless, it can be useful to consider superpositions containing negative $A_i$ to calculate amplitude spectra of fitness landscapes constructed by different means (see section \ref{rmfmodel} for an example).
   
%   Note that such a decomposition with arbitrary values for the $A_j$ can always be helpful for the computation of
%   spectra, but interpreting the system as a superposition of $NK$ landscapes in a biological sense can only reasonable 
%   if all the weights are positive.

  Interestingly, expression (\ref{eq:lkcompo}) is also obtained from another type of generalized
$NK$-model, giving rise to a different
  biological interpretation of the decomposition. Consider again fitness values $F(\sigma)$ that are constructed as sums of fitnesses corresponding to HoC-landscapes associated to $NK$-like neighborhoods
  $f_i(\sigma_{i_1},\dots,\sigma_{i_{k^{(i)}}})$,
    \begin{equation}
%         F:\mathbb H_2^N &\to \mathbb R\nonumber\\
        F: \sigma \mapsto \sum_{i=1}^M f_i(\sigma_{i_1},\dots,\sigma_{i_{k^{(i)}}}),\label{eq:variableneigh}
    \end{equation}
    where $M$ is an integer that can be different from $N$, and $k^{(i)}$ is the size of the $i$--th
    neighborhood, drawn from some distribution $P(k)$. Furthermore, for simplicity assume that the variances $D_i$ of the $f_i$ are all the same. The reasoning behind this model is to retain the idea of interacting groups of loci that is inherent in the $NK$-model, but to relax the rather unrealistic condition that all these groups are of the same size. Rather, it is assumed that there exist some typical distribution for the sizes of the groups. 

    Following the procedure explained in \cite{Campos2002}, the corresponding autocorrelation function is easily shown to be 
    \begin{align}
        R^{P}_d =\sum_{k\geq 0}P(k)\binom{N-k}{d} \binom{N}{d}^{-1}\label{linearlkmodel2},
    \end{align}
    which trivially leads to expression (\ref{eq:lkcompo}) with $A_k=P(k)$. The coefficients obtained from the decomposition of experimentally obtained spectra in terms of $NK$-spectra could therefore also be interpreted as a probability distribution for the sizes of interacting neighborhoods. Again, this interpretation is only consistent if all weights are positive. Here, it seems reasonable to expect that for large enough landscapes $P(k)$ should become continuous in the sense that the distribution becomes monotonic over large contiguous parts of its support.
% But unlike in the case were one tries to interpret the landscapes as superpositions of real $NK$ landscapes, here, one does not expect the decomposition to be sparse but, probably,  to be continuous.  

\section{Rough Mount Fuji model}
\label{rmfmodel}
  
Another model with tunable epistatic effects is the \textit{Rough Mount Fuji} (RMF) model \cite{aita}, which is constructed by superimposing a purely additive model and a HoC-landscape according to
\begin{equation}
%     F:\mathbb H_2^N &\to \mathbb R\nonumber\\
    F: \sigma \mapsto f_0+\sum_{i=1}^N b_{i}\sigma_i+ \xi(\sigma). \label{eq:rmf1def}
\end{equation}
In ref.~\cite{aita}, $f_0$ and the $b_i$ were parameters to be
determined empirically from experimental data. Here we instead choose
$f_0$ as some arbitrary constant, the $b_i$ as $N$ i.i.d. random variables, and $\xi(\sigma)$ as another
set of $2^N$ i.i.d. random variables with
$\langle\xi(\sigma)\rangle=0$ and
$\langle\xi(\sigma)\xi({\sigma^\prime})\rangle=D_N\delta_{\sigma\sigma^\prime}$,
compare to the construction of the HoC-model above in
sect.~\ref{nkmodel}. Note that, in contrast to the $\xi$, the $b_i$ do not
depend on $\sigma$. The amount of ruggedness is controlled by fixing
the variance of the HoC-component, $D_N$, and the mean of the absolute
values of the slopes of the additive model,
$s=\sum_{i=1}^N|b_i|/N$. The important limiting cases, the HoC-model
and the purely additive model, are obtained in the limits
$D_N/s\rightarrow \infty$ and $D_N/s\rightarrow 0$, respectively \cite{Szendro2012}.

 \begin{figure}[htb]

    \begin{center}
    \begin{minipage}{0.9\columnwidth}
    \includegraphics[width=\columnwidth]{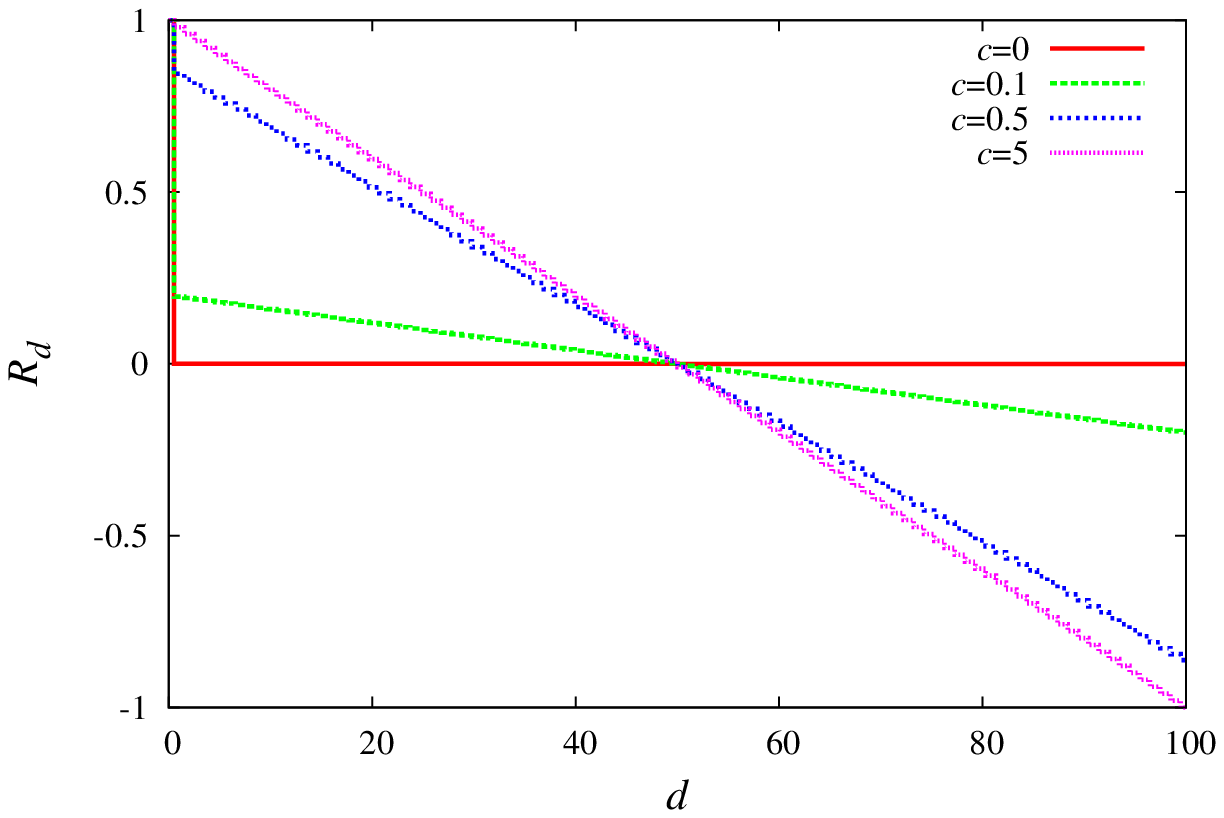}
    \end{minipage}
    \begin{minipage}{0.9\columnwidth}
    \includegraphics[width=\columnwidth]{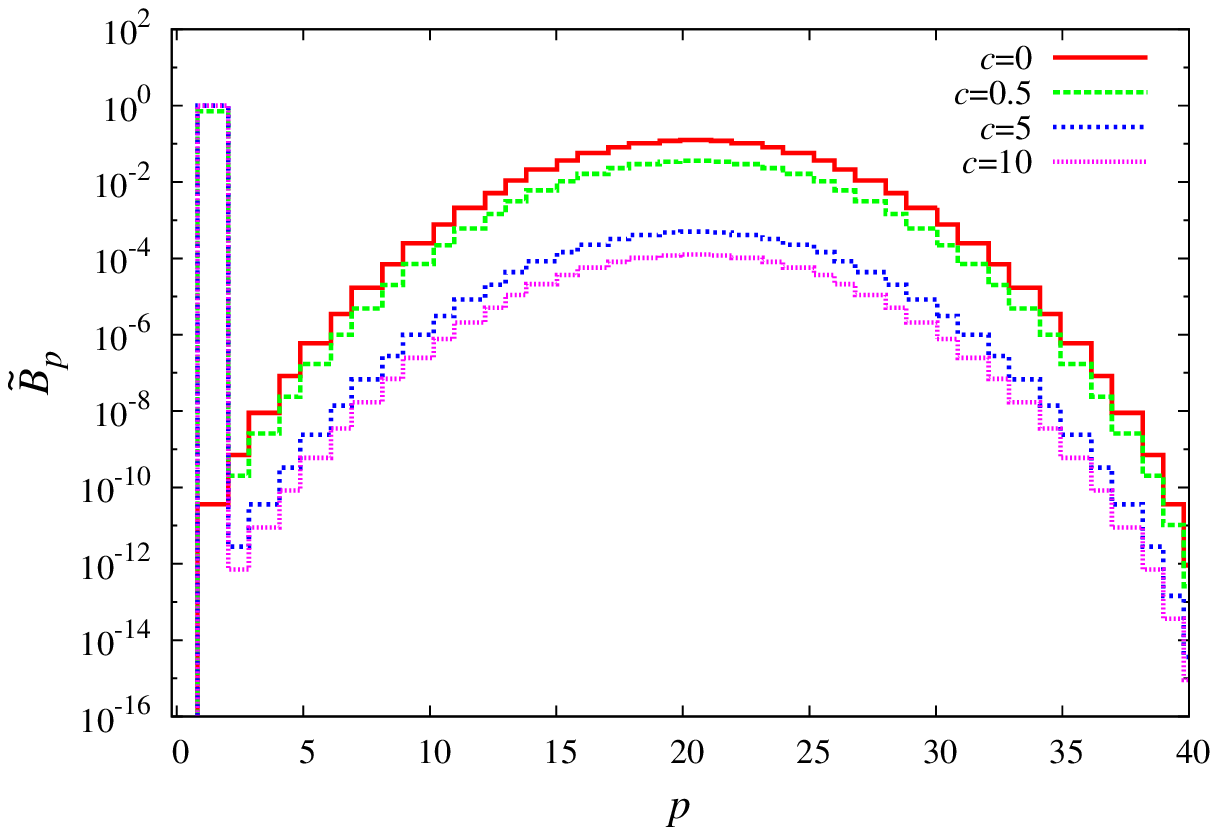}
    \end{minipage}
    \caption{The autocorrelation function (top) and the % normalized 
amplitude spectrum (bottom) for the RMF-model with $N=100$, $D_1=0$, $D_N = 1$ and  various values of $c$.}
    \label{fig:autocorrbprmf}
    \end{center}
    \end{figure}

In the following we write $b_i=\frac{c}{2}+\zeta_i$, where $c$ is a
constant independent of $i$, and the
$\zeta_i$ are i.i.d. random variables with $\langle\zeta_i\rangle=0$ and $\langle\zeta_i\zeta_j\rangle=D_1\delta_{ij}$. Note
that choosing the same mean value for all the $b_i$'s singles out the
\textit{reference sequence} $\sigma^{(0)}=(1,\ldots,1)$. On average, the fitness of
sequence $\sigma$ decays linearly with the Hamming distance
$d(\sigma,\sigma^{(0)})$ to the reference sequence $\sigma^{(0)}$ and
the mean slope is $c$.
Setting $D_1=0$ yields a simpler version of the RMF-model that was introduced in \cite{jaspjoa}.

To calculate the autocorrelation function of the RMF-model, it is convenient to rewrite the fitness as $F(\sigma)=\alpha-cd(\sigma,\sigma^{(0)})+\sum_{i=1}^N \zeta_{i}\sigma_i+ \xi(\sigma)$, where $\alpha=f_0+\frac{Nc}{2}$. Making use of the vanishing mean values of the $\zeta_i$'s and the $\xi$'s, 
the autocorrelation function reads
\begin{align*}
    R_d=&
    \Bigg(\left\langle\sum_{i=1}^N\zeta_i\sigma_i\sum_{j=1}^N\zeta_j\sigma^\prime_j\right\rangle_d+\langle\xi(\sigma)\xi(\sigma^\prime)\rangle_d\\
    &+\langle(\alpha-cd(\sigma,\sigma^{(0)}))(\alpha-cd(\sigma^\prime,\sigma^{(0)}))\rangle_d\\
    &-\left(\overline{\alpha-cd(\sigma,\sigma^{(0)})}\right)^2\Bigg)\cdot\\
    &\Bigg(\overline{(\alpha-cd(\sigma,\sigma^{(0)}))^2}-\left(\overline{\alpha-cd(\sigma,\sigma^{(0)})}\right)^2\\
    &+\left\langle\Bigg(\sum_{i=1}^N\zeta_i\sigma_i\Bigg)^2\right\rangle+\left\langle\xi(\sigma)^2\right\rangle\Bigg)^{-1}.
\end{align*}
The covariance of the deterministic part has been evaluated elsewhere\footnote{Neidhart, J., Szendro, I.G. \& Krug, J. Adaptation in tunably rugged fitness landscapes: The rough
Mount Fuji model (manuscript in preparation).} and the terms containing random variables can
easily be calculated, yielding
\begin{align}
    R^{\mathrm{RMF}}_d=\frac{\left(D_1+\frac{c^2}{4}\right)(N-2d)+D_N\delta_{d0}}{\left(D_1+\frac{c^2}{4}\right)N+D_N}.\nonumber
\end{align}

In order to obtain the spectrum $\tilde B_p$ we write $R^{\mathrm{RMF}}_d$ as a linear combination of correlation 
functions of the $NK$-model with different $k$'s, i.e.\
$R^{\mathrm{RMF}}_d=\sum_{k=0}^N A_k \binom{N-k}{d}/\binom{N}{d}$ with expansion coefficients
\begin{align} 
A_0&=-\frac{\left(D_1+\frac{c^2}{4}\right)N}{\left(D_1+\frac{c^2}{4}\right)N+D_N},
\; A_1=\frac{2\left(D_1+\frac{c^2}{4}\right)N}{\left(D_1+\frac{c^2}{4}\right)N+D_N},\nonumber\\ 
A_N&=\frac{D_N}{\left(D_1+\frac{c^2}{4}\right)N+D_N}, 
\end{align} 
and $A_k=0$ for all other $k$'s. The $\tilde B_p$ can now be calculated
making use of the linearity of equation (\ref{eq:bq}), yielding
\begin{align}
 \tilde B^{\mathrm{RMF}}_p=\frac{\left(D_1+\frac{c^2}{4}\right)N \delta_{p1}+D_N 2^{-N}\binom{N}{p}}{\left(D_1+\frac{c^2}{4}\right)N+D_N}.
\end{align}
% Where $\binom 0 p = \delta_{p,0}$. 
In fig.\ \ref{fig:autocorrbprmf}, autocorrelation functions and
amplitude spectra for the RMF-model with $D_1=0$ and various choices
of $c$ are shown. Note that the generality of the superposition ansatz
made it possible to calculate the $\tilde B_p$ for the RMF-model,
although the relation to the $NK$-model is not obvious at first
sight. Having in mind that the zeroth component does not contain
information about epistasis, we adopt, for the rest of this paper, a
more general definition of RMF-landscapes as superpositions of $NK$
-landscapes with all components being equal to zero, except for
$A_1>0$, $A_N>0$, and an arbitrary zeroth order coefficient $A_0$ that may be of any sign.

\section{Exponentially decaying correlation functions}
\label{exponentialcorr}
\begin{figure}
\includegraphics[width=0.9\columnwidth]{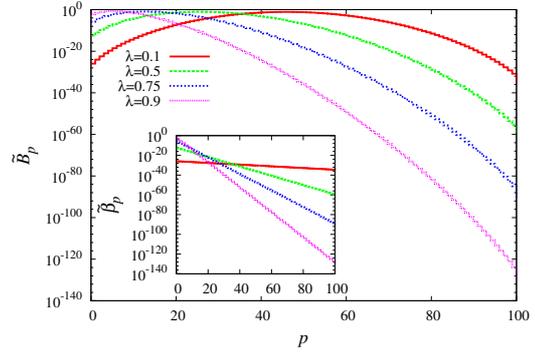} %bp_exp.eps}
\caption{The % normalized 
amplitude spectrum $\tilde B_p$ (main) and the renormalized spectrum $\tilde\beta_p$ (inset) for
exponentially decaying fitness correlations. In the inset, the exponential decay is obvious.}
\label{fig:bpexp}
\end{figure}

The motivation for the present paper is to identify typical features
of amplitude spectra of fitness landscapes and to make
use of them for extracting information  about the underlying biological system. In the preceding  two
sections we considered well-established statistical models of fitness
landscapes and computed their spectra. As will be further illustrated
in sect.~\ref{experimentalfl}, this analysis provides criteria to judge whether a measured spectrum can
be explained by these models or not and, if so, one can use the biological picture behind the model to try to interpret the findings. 

However, when faced with experimental data, none of the presented models may be general enough to give a good
description. If this is the case, an alternative ansatz is to start
with a  presumably generic correlation function and calculate the corresponding spectrum, which can be compared to the data.
This may then also guide the search for improved models. 
Here, we consider a correlation function that decays exponentially with Hamming distance $d$
\begin{equation}
R^\mathrm{exp}_d=\lambda^d, \label{eq:correexp}
\end{equation}
with $0 < \lambda < 1$. 
The resulting expression for the spectrum obtained from eq.\ (\ref{eq:bq}), 
\begin{equation}
\label{Bqexp}
\tilde B_q = 2^{-N} \sum_{d \geq 0} \binom{N}{d} K^{(2)}_q(d) \lambda^d,
\end{equation}
is most easily evaluated using the known form of the generating
function of the Krawtchouk polynomials \cite{Szego,stoll}
\begin{equation}
\label{K_gen}
{\cal K}^{(2)}(x,z) = \sum_{n \geq 0} K^{(2)}_n(x) \, z^n = (1 - z)^x (1+z)^{N-x}
\end{equation}
and the fact that these polynomials are self-dual in the sense of \cite{Koekoek2010}
\begin{equation}
\label{self-dual}
\binom{N}{x} K^{(2)}_n(x) = \binom{N}{n} K^{(2)}_x(n).
\end{equation}
Indeed, inserting (\ref{self-dual}) into (\ref{Bqexp}) and using
(\ref{K_gen}) yields 
\begin{equation}
\label{Bqexp_final}
\tilde B_q = 2^{-N} \binom{N}{q} (1 - \lambda)^q (1 + \lambda)^{N-q}.
\end{equation}
Defining $\kappa=\ln{\left(\frac{1+\lambda}{1-\lambda}\right)}$ this expression can be rewritten as 
\begin{align}
\tilde B_q=\frac{\binom{N}{q}}{(1+\mathrm{e}^{-\kappa})^N}\mathrm{e}^{-\kappa q}, \label{eq:bexp}
\end{align}
corresponding to
$R^\mathrm{exp}_d=\left(1-\frac{2}{1+\mathrm{e}^{\kappa}}\right)^d$. We
conclude that if the spectrum normalized with respect to
the number of $q$--tuples, $\tilde \beta_q=\tilde B_q/\binom{N}{q}$,
decays exponentially with $q$, then the correlations decay
exponentially with distance on the hypercube, see fig. \ref{fig:bpexp}. 

Although we are, at the moment, lacking simple stochastic models that
produce exponentially decaying correlations, spectra of the form
(\ref{eq:bexp}) have recently been found for fitness landscapes
obtained from a dynamical model of molecular signal transduction
\cite{Pumir2011}. It would be interesting to see whether one
can construct stochastic models that do not enter too deeply into the 
dynamics at the cellular level but contain a simple and
generic mechanism that gives rise to such correlations.

Exponentially decaying correlations have also been reported in a
recent large-scale study of the fitness landscape of HIV-1
\cite{Kouyos2012}. However, the correlation function calculated in
that article is different from the one studied here, as it averages
over correlations between fitness values of mutants that are connected
by random walks of some length $s$ and not over fitness values
corresponding to states separated by Hamming distance $d$. Such random
walk correlation functions are also connected in a simple manner to
the amplitude spectra \cite{stadler1}, but the relation is different
from the one considered here. Therefore our results are not directly applicable to these observations.

\section{Experimentally obtained fitness landscapes}
\label{experimentalfl}

 \begin{figure*}
    \begin{center}
    \includegraphics[width=.7\textwidth]{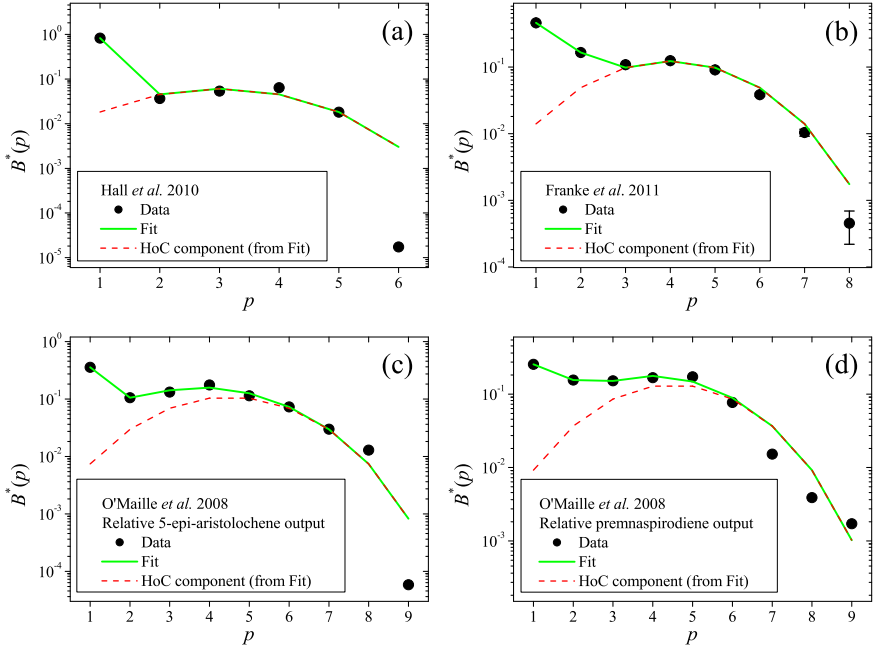}
    \caption{Spectra corresponding to various experimentally measures fitness landscapes. The green lines are obtained by fitting the spectrum of a superposition of $NK$-models to the data. The dashed red line is proportional to $\binom{N}{p}$, showing the spectrum expected for a HoC-component.}
    \label{fig:empirical}
    \end{center}
    \end{figure*}

In this section we compare the model spectra to several
experimentally measured ``fitness'' landscapes. The quotation marks
indicate that not all of the landscapes presented here actually
correspond to fitness, but rather to some proxy of it. To be able to
compare spectra, the landscapes should be as large and as complete as
possible. The four landscapes considered are a six locus landscapes
obtained by Hall {\it et al}.\ \cite{Hall2010} for the yeast
\textit{Saccharomyces cerevisiae}, an eight locus landscape for the
fungus \textit{Aspergillus niger} presented by Franke {\it et al}.\
\cite{jaspjoa}, and two nine locus landscapes for the plant \textit{Nicotiana
tabacum} studied by O'Maille {\it et al}.\ \cite{OMaille2008}. A
comparative analysis of these (and other) empirical landscapes can be
found in \cite{Szendro2012}. 
All spectra presented in this section were calculated directly by decomposing the fitness landscapes in terms of the eigenfunctions of the graph Laplacian.

While the first two landscapes mentioned above measure
growth rate as a quantifier of fitness, the landscapes presented in \cite{OMaille2008} measure
enzymatic specificity of terpene synthases, that is, the relative production of 5-epi-aristolochene and premnaspirodiene,
respectively. As for these landscapes only 418 out of 512 fitness values were measured, the missing data is estimated
by fitting a multidimensional linear model \cite{Aita2001} to the
measured landscape. The fitness values of states for which there are
no measurements are then replaced by the values given by the linear
model. On the contrary, for the \textit{A. niger} landscape considered in \cite{jaspjoa},
missing fitness values were argued to correspond to non-viable mutants and are therefore set to zero. The way of estimating missing values obviously affects the spectra, but some estimation is necessary to be able to carry out the analysis. 

We now ask whether the experimental spectra can be expressed as
superpositions of $NK$-spectra of the form (\ref{eq:lkcompo}) (recall that the RMF-model is a particular case of such a superposition). Of course,
such a decomposition is always possible, but the assumption that the biological mechanism responsible for the spectra is
really the additive interplay of fixed groups of loci of characteristic sizes is only reasonable if all the coefficients
$A_j$ are positive.

Simply solving the linear system of equations (\ref{eq:lkcompo})
generally yields several negative coefficients. More satisfactory results are obtained by fitting a function of the form (\ref{eq:lkcompo}) to the data by means of a
least square procedure, constraining the coefficients to positive
values. Here, two ansatzes are considered. First a fit containing all
coefficients is carried out, with none of the $A_j$ fixed to zero
\textit{a priori}. This is done to check whether a superposition of
type (\ref{linearlkmodel2}) with a continuous neighborhood size
distribution $P(k)$ is appropriate. Second, sparse fits containing as
few nonzero $A_j$'s as possible are carried out to verify if the
landscape could be biologically interpreted as a superposition of a
small number of $NK$-landscapes of different interaction ranges. One
way of selecting $A_j$'s that can be neglected in the fit
is to identify those coefficients obtained in the full fit that are
much smaller than the others. 
In all cases, the term proportional to $A_0$ in (\ref{eq:lkcompo}) is not considered as it can always be trivially fixed to fit $\tilde B_0$.   

In fig.\ \ref{fig:empirical} the data for the normalized amplitudes $B_p^\ast$ (black dots) is shown together with the fit (green curve)
and the HoC-component  $\sim \binom{N}{p}$ (red dashed line) of the fit.
For the \textit{A. niger} landscape in \cite{jaspjoa} error estimates for the
fitness values were available \cite{Franke_Diss}, enabling the
calculation of error bars to the spectrum. This is done by
constructing ensembles of landscapes with fitness values $F(\sigma)=
\langle F(\sigma) \rangle +\xi(\sigma)$, where $\langle F(\sigma)
\rangle$ is the mean of the replicate experimental measurements of the fitness of genotype $\sigma$ and
the $\xi(\sigma)$ are normally distributed random numbers with $\sigma$-dependent
standard deviations obtained from the replicate measurements. 
Note that the influence of the measurement errors on the
spectra is very small and only exceeds the symbol size for the highest
$p$ component ($p=8$). At least for this case one can therefore
safely exclude that the HoC-component of the spectrum is generated by measurement errors.

As can be seen in fig.\ \ref{fig:empirical}(a), the spectrum of the
yeast landscape \cite{Hall2010} is nicely fitted by an ansatz where
only $A_1$ and $A_N$ are assumed to be different from zero. This is
evidently a superposition of an additive and a HoC-landscape and therefore a
RMF-landscape. Only the value at $p=N$ seems too small to be fitted by
the model. However, this value corresponds to a single component of
the decomposition (\ref{eq:ft}) and the large deviation may be due to
the lack of averaging. Also for the \textit{A. niger} 
landscape from \cite{jaspjoa} a nice and sparse fit with nonzero
coefficients $A_1$, $A_2$, and $A_N$ is obtained (see Fig.\
\ref{fig:empirical}(b)). The significant value of $A_2$ implies that
there are important interactions between pairs of loci. A RMF
-landscape is therefore not an appropriate model of this system. Note
that this conclusion differs from the analysis presented in
\cite{jaspjoa}, where a reasonable fit to the RMF-model was found
for a particular epistasis measure, the number of accessible pathways.
This illustrates the importance of using more than one topographic
measure for the comparison between empirical and model landscapes
\cite{Szendro2012}.   
% However, at this point, it cannot
% be decided whether the pair interactions do really only take place
% between exactly $N$ fixed pairs of loci, as is suggested by the $NK$
% picture. 
 
%In both cases, considering the full ansatz does not yield a qualitative improvement of the fit and provides no evidence for an underlying %continuous neighborhood size distribution $P(k)$. 

For the spectrum of the 5-epi-aristolochene \textit{N. tabacum} landscape from \cite{OMaille2008}, the fitting yields reasonable results for an ansatz allowing only $A_1$, $A_2$, $A_6$ and $A_N$ to be different from 0 (see Fig.\ \ref{fig:empirical}(c)). This might indicate that, apart from the non epistatic part and the simple pair interactions, there are one or several groups consisting of six strongly interacting alleles. Using the same ansatz for the premnaspirodiene landscape yields less convincing results, as the large $p$ part of the spectrum seems to be poorly fitted (see Fig.\ \ref{fig:empirical}(d)). Introducing more components into the fitting ansatz yields better results for this part of the spectrum, but such ansatzes can hardly be considered sparse anymore. 

Using the full ansatz to fit the different landscapes does not yield
any qualitative improvement for the first three landscapes and
provides no evidence for an underlying continuous neighborhood size
distribution $P(k)$. Only for the premnaspirodiene
\textit{N. tabacum} landscape does the fit for the spectrum improve
notably, but the obtained spectrum does not support the idea of a
continuous distribution of neighborhood sizes (not shown). In general,
such a continuous distribution is more likely to emerge for larger
landscapes than the relatively small data sets considered here, which
suffer from insufficient averaging over groups of loci of different
sizes. 

One should be aware that failing to obtain a reasonable decomposition
of an empirical landscape in terms of $NK$-spectra does not
\textit{a priori} rule out the possibility that the landscape is
in fact shaped by the mechanisms assumed by a superposition of
$NK$-models. For example, the failure may be due to an
\textit{inappropriate} fitness measure, in the following
sense. Suppose that there exists a fitness proxy, $F^\prime(\sigma)$,
whose decomposition in terms of $NK$-landscapes is sparse, but the
proxy actually measured in experiments is
$F=G(F^\prime)$, with $G$ being some nonlinear
function. The decomposition of $F$ may then not be sparse anymore and the biological mechanism that shapes the landscape may be obscured. 

Finally, it was checked whether any of the spectra are compatible with
the expression  (\ref{eq:bexp}) corresponding to an exponentially
decaying correlation function, but no reasonable
correspondence was found. Of course, this does not allow for the conclusion that exponentially decaying correlations are an
unrealistic assumption. Possibly, it may again be necessary to go to larger landscape sizes to see such behavior. Also, the
way in which the mutations constituting the landscape are selected may have an influence on the observed correlations (see e.g.\ \cite{Szendro2012}).

\section{Conclusions}
\label{conclusions}

Exploiting the connection between amplitude spectra and fitness autocorrelation functions of fitness
landscapes over the Boolean hypercube, the amplitude spectrum of
Kauffman's $NK$-model
was calculated exactly and found to be of the simple form (\ref{eq:lkspectra}).
By superimposing $NK$-landscapes the spectra of RMF-type models could also be obtained.
In addition, an $NK$-like model with a distribution $P(k)$ of neighborhood sizes was introduced and its spectrum was calculated.
Such an extension of the $NK$-model is reasonable, because it cannot be assumed in general that every locus interacts
with the same number of other loci. This model thus offers more flexibility to fit
experimental data. As a last example, the spectrum of a model with exponentially decaying correlations was computed.

The HoC, RMF and $NK$-models are frequently used for analyzing evolutionary
processes, classifying fitness landscape properties and fitting experimental data. Therefore a lot of effort has been invested in the
understanding of these models, but the link to experimental data is
still rather weak. The amplitude spectra calculated in this article
should facilitate quantitative comparisons in future studies. The
spectra contain a large amount of information about the landscape
topography, and it is important to understand how the spectrum encrypts this
information in order to be able to interpret the spectra of measured fitness landscapes.
As an exemplary application of our results, four experimental landscapes were fitted by means of the model spectra. Three of them could be fitted very nicely with sparse superpositions of $NK$-models, while for the fourth one the obtained fit seems less convincing. In none of the cases evidence for a continuous neighborhood size distribution $P(k)$ was found, which might be due to the small sizes of the landscapes discussed in this article.    

We claim that the fitting of amplitude spectra can be a useful tool
for data analysis, but it has to be emphasized that the spectra cannot
be assigned to model landscapes in a unique way. Also, the collection
of models presented here is by no means exhaustive. Obtaining
analytical expressions for the amplitude spectra of other classes of
fitness landscapes is desirable and should prove helpful in guiding the search for suitable models of experimental landscapes.

Finally, it is important to mention that there are interesting and
biologically relevant properties of fitness landscapes that cannot be
obtained from their spectra, such as, for example, the number of local
fitness maxima and the number of selectively accessible pathways
\cite{Weinreich2006,jaspjoa}. While it was shown in ref.~\cite{Szendro2012}
that the ruggedness measure $B_\mathrm{sum}$ based on the Fourier
decomposition correlates with both quantities, there is no strict
correspondence between these measures of epistatic
interactions. Amplitude spectra do not distinguish
between different kinds of epistasis, i.e.\ magnitude, sign, or
reciprocal sign epistasis, in a qualitative way. 
Therefore, if one is interested in this distinction, other epistasis
measures have to be included in the analysis.

\section*{Acknowledgments}

We thank B. Schmiegelt, P.F.\ Stadler and D.M. Weinreich for useful
discussions and correspondence, and D. Hall for providing the original
data of the \textit{S. cerevisiae} landscape.
This work was supported by DFG within SFB 680, SFB-TR 12, SPP 1590 and the Bonn Cologne Graduate School for Physics and Astronomy.

\appendix

\section{Fourier spectrum of the $NK$-model}
\label{Appendix_NK}
    To evaluate the expression (\ref{Bq_NK}), an alternative but equivalent formulation for
    the Krawtchouk polynomials is needed. With \cite{stoll} 
$$K^{(2)}_q(d)=\sum_{i\geq 0}(-2)^i\binom d i \binom{N-i}{q-i}$$
we obtain
    \begin{align*}
        \tilde B_q&=\sum_{d\geq 0}K^{(2)}_q(d)\binom{N-k}{d}\nonumber\\
        &=2^{-N}\sum_{i\geq 0}\sum_{d\geq 0}(-2)^i \binom d i \binom{N-i}{q-i}\binom{N-k}{d}.
    \end{align*}
The summation over $d$ can be carried out using the identity \cite{gould} 
$$\sum_{d\geq 0} \binom d i \binom {N-k} d = 2^{N-k-i}\binom {N-k} i,$$ which yields
    \begin{equation} \label{Bq_sum2}
        \tilde B_q
%         &=2^{-N}\sum_{i\geq 0} (-2)^i\binom{N-i}{q-i}\binom{N-k}{i} 2^{N-k-i}\nonumber\\
	=2^{-k}\sum_{i\geq 0}(-1)^i\binom{N-i}{q-i}\binom{N-k}{i}.
    \end{equation}
At this point we relax the condition (\ref{convention}) of positivity on the entries of the binomial
coefficients. This allows us to perform an `upper negation' \cite{concrete_math} in the first binomial factors in eq.(\ref{Bq_sum2}), 
$$\binom{N-i}{q-i} = (-1)^{q-i} \binom{q-N-1}{q-i}.$$
The remaining sum over $i$ can now be evaluated using the Vandermonde identity \cite{concrete_math},
\begin{align*}
\tilde B_q &= 2^{-k} (-1)^q \sum_{i\geq 0 }\binom{q-N-1}{q-i}
    \binom{N-k}{i}\nonumber\\
&=2^{-k} (-1)^q \binom{q-k-1}{q}
\end{align*} 
and with another upper negation we arrive at the final result (\ref{eq:lkspectra}).

%% References

\bibliographystyle{model1a-num-names}

\end{document}